\documentclass[prl,twocolumn,showpacs]{revtex4-1}

\newcommand{\cP}{{\cal P}}

\newcommand{\Eq}[1]{Eq.~(\ref{eq:#1})}

\newcommand{\f}{\mathbf{f}}
\newcommand{\Fig}[1]{Fig.~\ref{fig:#1}}

\newcommand{\REF}[1]{Ref.~\cite{#1}}

\renewcommand{\f}{\mathbf{f}}

\newcommand{\rr}{\mathbf{r}}
\renewcommand{\v}{\mathbf{v}}

\newcommand{\expt}[1]{\left< #1\right>}
\newcommand{\gdot}{\dot{\gamma}}

\usepackage{graphicx}
\usepackage{amssymb}
\usepackage{amsmath}
\usepackage{bm}

\begin{document}
\title{Slow and fast particles in shear-driven jamming: critical behavior and finite size scaling}

\author{Peter Olsson}

\affiliation{Department of Physics, Ume\aa\ University, 
  901 87 Ume\aa, Sweden}

\date{\today}   

\begin{abstract}
  We do shear-driven simulations of a simple model of non-Brownian particles in two
  dimensions. By examining the velocity distribution at different densities and shear
  rates we find strong evidence for the existence of two different processes, respectively
  dominated by the slower and the faster particles---the slow process and the fast
  process. The leading divergence in the shear viscosity is governed by the fast process.
  An examination of height and position of the low-velocity peak in the distribution
  demonstrates that it is the slow process that is responsible for the
  correction-to-scaling term in the critical scaling analysis. We further find that the
  presence of velocity correlations across large distances is primarily due to the slow
  process which implies that the diverging viscosity and the diverging correlation length
  are only indirectly related.
\end{abstract}
\maketitle
Particle transport is ubiquitus in both industry and every-day life with varying behaviors
due e.g.\ to differing particle properties and geometries. A reasonable approach in the
quest for a better understanding of the slowing down of the dynamics, e.g.\ because of an
increase in density, is to first examine simplified models. One of the simplest possible
consists of a collection of circular disks in two dimensions with contact-only
interactions \cite{OHern_Silbert_Liu_Nagel:2003}. When such a collection of particles is
driven at a constant shear strain rate $\gdot$ \cite{Durian:1995}, the system develops a
shear stress $\sigma$, and the shear viscosity, $\eta\equiv\sigma/\gdot$, diverges as the
jamming density, $\phi_J$, is approached from below. For the conceptually simple case of
hard disks and overdamped dynamics this is seen in an algebraic divergence,
$\eta \sim (\phi_J-\phi)^{-\beta}$. This divergence is related to the increase in contact
number $z$ towards the isostatic value, $z_c-z\sim(\phi_J-\phi)^{u_z}$.  In the
thermodynamic limit the isostatic contact number is $z_c=2d$ \cite{Alexander:1998}; see
\REF{Goodrich:2012} for the generalization to finite $N$. In terms of the contact number
deficit the divergence becomes $\eta\sim(z_c-z)^{-\beta/u_z}$.

A hallmark of critical phenomena is a diverging spatial correlation length. It has also
long been realized that particle motion in sheared systems becomes increasingly collective
as the jamming density is approached \cite{Pouliquen:2004, Lechenault_2008,
  Heussinger_Berthier_Barrat:2010, Hexner:2018}. In a recent analysis of the velocity field
\cite{Olsson_Teitel:jam-xi-ell} it was further found that fluctuations in the rotation and
the divergence of the velocity field behave differently, and that the length related to
the rotations appears to be the more important one, diverging as
$\xi\sim(\phi_J-\phi)^{-\nu}$, with $\nu\approx 1$ \cite{Olsson_Teitel:jam-xi-ell}.

\begin{figure}
  \includegraphics[bb=51 324 314 659, width=4.2cm]{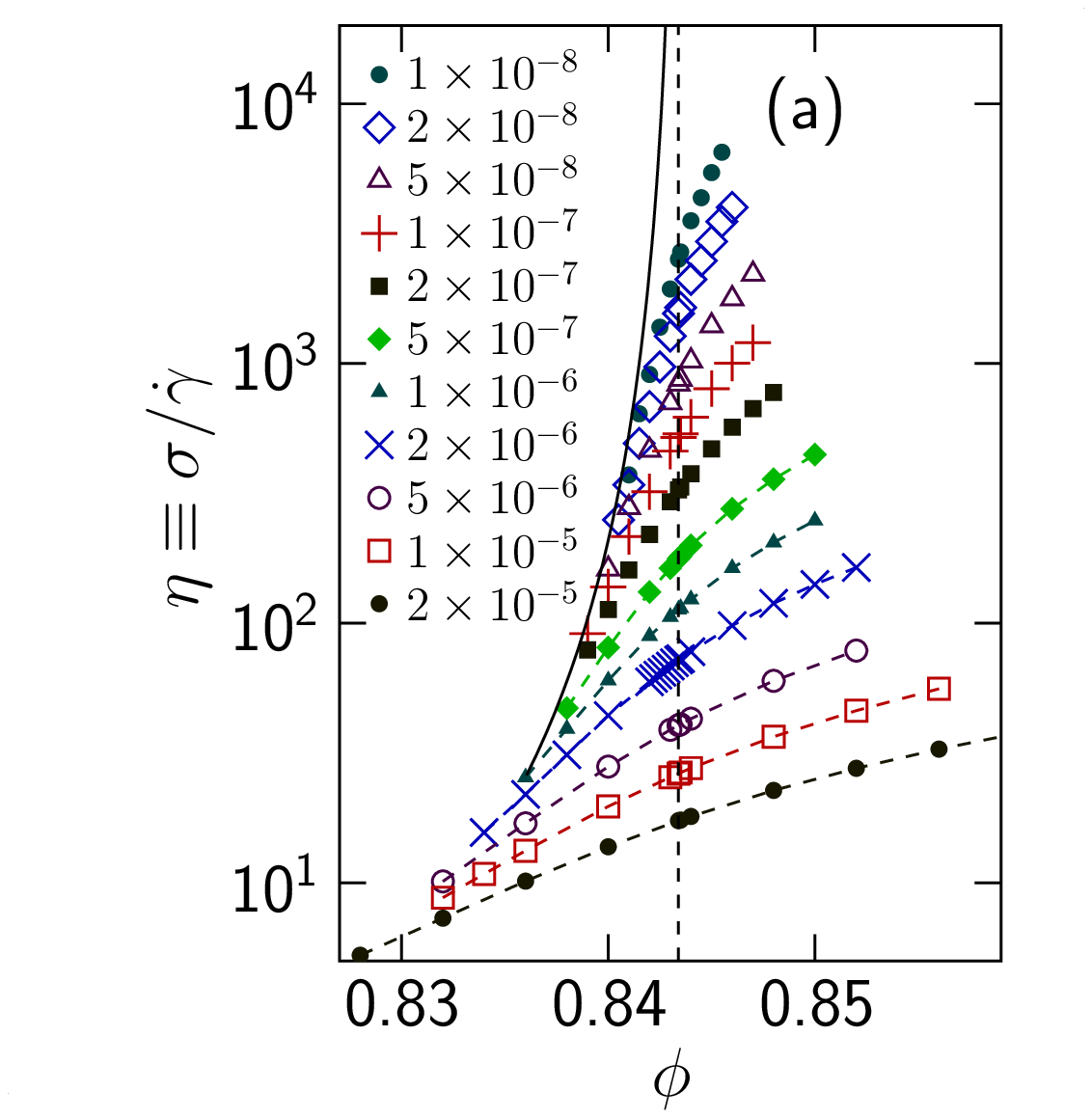}
  \includegraphics[bb=51 324 314 659, width=4.2cm]{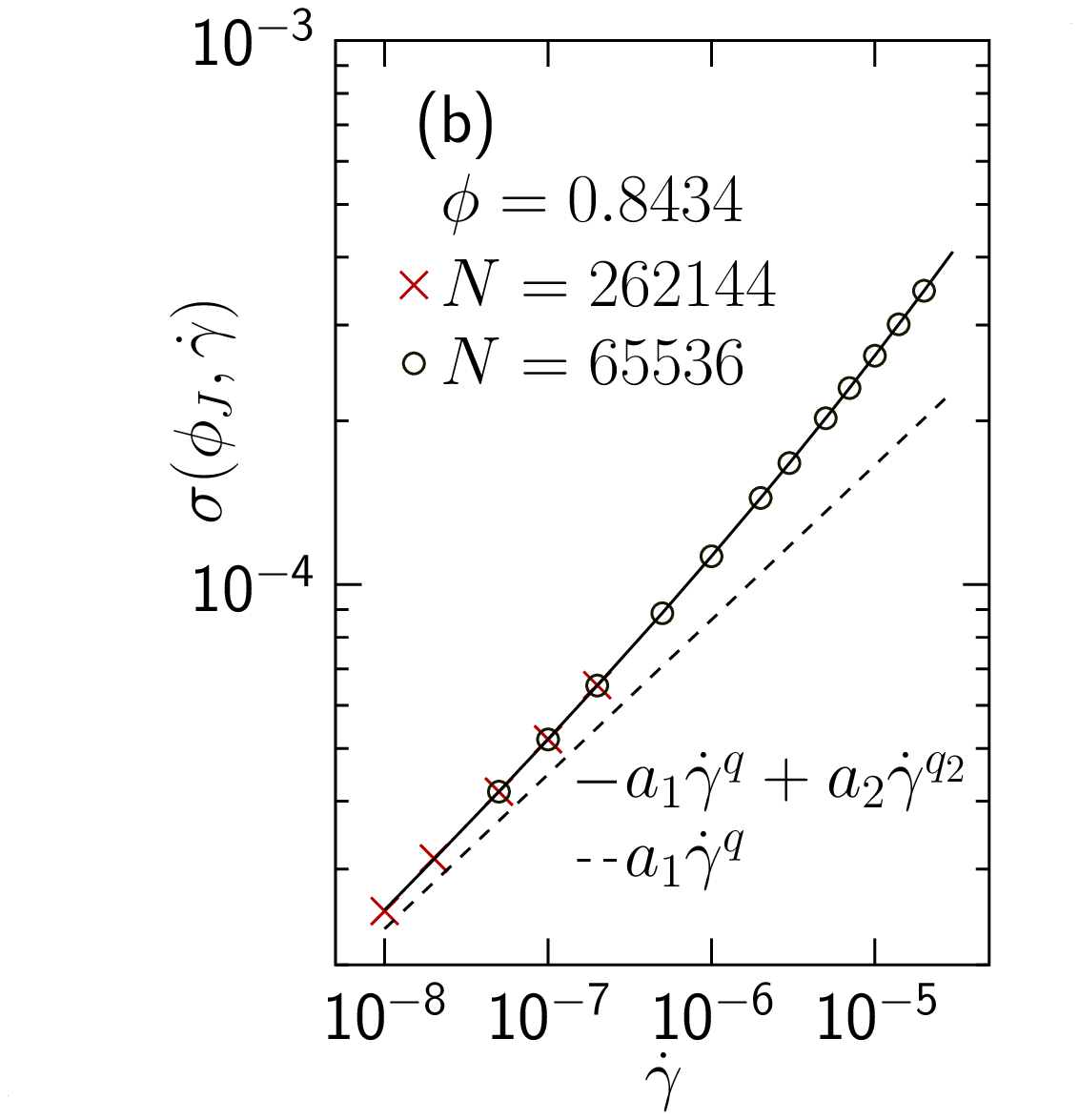}
  \caption{The divergence of the shear viscosity. Panel (a) is the shear viscosity
    $\eta\equiv\sigma/\gdot$ for different shear strain rates. The vertical dashed line is
    $\phi_J$ and the solid line is the approximate critical divergence
    $\eta(\phi,\gdot\to0) \sim(\phi_J-\phi)^{-\beta}$, $\beta=2.7$. Panel (b) which is
    $\sigma$ vs $\gdot$ at $\phi=0.8434\approx\phi_J$ illustrates the determination of the
    exponents $q$ and $q_2$ defined in \Eq{sigma-at-phiJ}.}
  \label{fig:sigma}
\end{figure}

Because of difficulties with numerically simulating the dynamics of hard particles,
simulations are commonly performed with elastic particles with forces related to the
particle overlaps. The shear viscosity has then a strong shear strain rate
dependence---see \Fig{sigma}(a)---and in attempted critical scaling analyses
\cite{Olsson_Teitel:jamming, Olsson_Teitel:gdot-scale} it has furthermore become clear
that one also needs to include a correction-to-scaling term
\cite{Olsson_Teitel:gdot-scale, Kawasaki_Berthier:2015}.  In two dimensions (2D) such
determinations of $\beta$ have typically given values in the range $\beta=2.2$ through
2.83 \cite{Andreotti:2012, Olsson_Teitel:gdot-scale, Kawasaki_Berthier:2015}. For the
combination $\beta/u_z$ different methods that directly probe the hard disk limit give
$\beta/u_z=1/0.38=2.63$ \cite{Lerner-PNAS:2012} and $\beta/u_z=2.69$
\cite{Olsson:jam-tau}. (Determinations in three dimensions tend to give higher values,
$\beta/u_z\approx 3.3$ \cite{DeGiuli:2015} or $\beta/u_z=3.7\pm0.7$
\cite{Olsson:jam-3D}. An attempt to explain this dependence on dimensionality in terms of
a finite size effect is discussed in \cite{Nishikawa_Ikeda_Berthier:2021,
  Olsson:jam-NIB}.)

From analytical considerations the exponent has however been argued to be
$\beta/u_z\approx3.41$ \cite{DeGiuli:2015, Harukuni-logcorr:2020} and it has then been
claimed that the determinations quoted above for 2D are incorrect due to the neglect of
logarithmic corrections to scaling \cite{DeGiuli:2015, Harukuni-logcorr:2020,
  Nishikawa_Ikeda_Berthier:2021}.  Though this explanation is a possibility, it could also
be that the discrepancy only points to a lack of understanding of the phenomenon of
shear-driven jamming.

In the present Letter we present evidence for a novel picture that describes shear-driven
jamming as being controlled by two different processes dominated by the slow and the fast
particles, respectively, and accordingly coined the ``slow process'' and the ``fast
process''. The fast process is responsible for the leading term in the divergence of the
shear viscosity and is dominated by particles in the tail of the velocity distribution
\cite{Olsson:jam-vhist}. The slow process leads to the correction-to-scaling term
\cite{Olsson_Teitel:gdot-scale, Kawasaki_Berthier:2015}, and arises from particles at and
below the low-velocity peak in the distribution. It is further found that the presence of
velocity correlations across large distances is related to the slow process whereas the
fast process appears to be short range correlated, only. The present Letter gives a short
description of a comprehensive examination of shear-driven jamming; a more detailed
discussion is given in \REF{jointPRE}, except for the finite size dependence, which will
be discussed elsewhere \cite{Olsson:slow-fast:fss}.  The analyses presented here work the
same also in three and four dimensions, presentations of these results will however also
be deferred to a later publication.

We simulate a bidisperse collection of particles in 2D with equal number of particles of
two different sizes. The total number of particles is $N=65536$ particles, if not
otherwise noted. The small particles have diameter $d_s=1$ and the size ratio is 1.4
\cite{OHern_Silbert_Liu_Nagel:2003}. For particles in contact we define the relative
overlap $\delta_{ij}=1-r_{ij}/d_{ij}$ where $r_{ij}$ is the distance between particles $i$
and $j$ and $d_{ij}$ is the sum of their radii. The contact interaction is from the
potential energy $V_p(r_{ij}) = \epsilon \delta^2_{ij}/2$; we take $\epsilon=1$. The force
on particle $i$ from particle $j$ is $\f^\mathrm{el}_{ij} = -\nabla_i V_p(r_{ij})$, which
gives $f^\mathrm{el}_{ij}=\epsilon\delta_{ij}/d_{ij}$. We do shearing simulations with a
time-dependent shear strain $\gamma=\gdot t$ and Lees-Edwards boundary conditions
\cite{Evans_Morriss} on a system with volume $V=L\times L$. The shearing gives an average
homogenous velocity profile $=\gdot y\hat x$ but our focus will be on the non-affine
velocity, which is the particle velocity relative to this velocity profile,
$\v_i=\v^\mathrm{tot}_i-\gdot y_i\hat x$. Related to the non-affine velocity is the
dissipative force $\f^\mathrm{dis}_i=-k_d\v_i$. We simulate with overdamped dynamics such
that $\f^\mathrm{el}_i + \f^\mathrm{dis}_i= 0$ which becomes
$\v_i = \f^\mathrm{el}_i/k_d$. We take $k_d=1$ and the unit of time
$\tau_0=d_s^2k_d/\epsilon=1$. The shear stress is
$\sigma = - \expt{\mathbf{p}^\mathrm{el}_{xy}}$ from the pressure tensor which is obtained
from the forces between the contacting particles,
$\mathbf{p}^\mathrm{el} = V^{-1}\sum_{i<j} \f^\mathrm{el}_{ij}\otimes\rr_{ij}$. Because of
the large $N$ in our simulations the fluctuations in pressure
$\equiv \frac 1 2(\mathbf{p}^\mathrm{el}_{xx}+\mathbf{p}^\mathrm{el}_{yy})$ during the
run are very small.

The analyses below will be done in terms of the shear stress $\sigma$ but to illustrate
the jamming transition \Fig{sigma}(a) shows
$\eta(\phi,\gdot)\equiv\sigma(\phi,\gdot)/\gdot$ for shear strain rates $\gdot=10^{-8}$
through $2\times10^{-5}$. The transition is shown by the rapid increase of $\eta$ with
$\phi$, which in the $\gdot\to0$ limit approaches $\eta\sim(\phi_J-\phi)^{-\beta}$, with
$\beta\approx2.7$, illustrated by the solid line. The analyses of this kind of data (and
the similar $\eta_p=p/\gdot$) in the literature \cite{Olsson_Teitel:jamming,
  Olsson_Teitel:gdot-scale} rely on the standard scaling assumption \cite{Vagberg_Olsson_Teitel:CDn},
\begin{equation}
  \label{eq:sigma-b}
  \sigma(\phi, \gdot)b^{y/\nu} =  \bar g_\sigma(\delta\phi\, b^{1/\nu}, \gdot b^z)
  + b^{-\omega} \bar h_\sigma(\delta\phi\, b^{1/\nu}, \gdot b^z).
\end{equation}
Here $b$ is a length rescaling factor, $y$ is the scaling dimension of $\sigma$, $\nu$ is
the correlation length exponent, $\delta\phi=\phi-\phi_J$, $z$ is the dynamical exponent,
$\omega$ is the correction-to-scaling exponent and $\bar g_\sigma$ and $\bar h_\sigma$
are unknown scaling functions. With $b=\gdot^{-1/z}$ in \Eq{sigma-b} together with
$q=y/z\nu$ and $q_2=q+\omega/z$ one finds
\begin{equation}
  \label{eq:sigma-scale}
  \sigma(\phi,\gdot) = \gdot^q g_\sigma\left(\frac{\phi-\phi_J}{\gdot^{1/z\nu}}\right)
    + \gdot^{q_2} h_\sigma\left(\frac{\phi-\phi_J}{\gdot^{1/z\nu}}\right).
\end{equation}
In the fitting of \REF{Olsson_Teitel:gdot-scale} the scaling functions were taken to be
exponentials of polynomials in $(\phi-\phi_J)/\gdot^{1/z\nu}$, and the parameters of these
polynomials together with $\phi_J$ and the critical exponents were adjusted to get the
best possible fit. [Since $\omega>0$, as it is an irrelevant scaling variable, it follows
that $q_2>q$, and that for $\eta$ the two terms of \Eq{sigma-scale} scale as $\gdot^{q-1}$
and $\gdot^{q_2-1}$, which implies that the first term is the more divergent one as
$\gdot\to0$.]

As shown in \Eq{sigma-scale} the scaling expression for $\sigma(\phi,\gdot)$ consists of
two different terms, as first reported in \REF{Olsson_Teitel:gdot-scale}, and the question
that we set out to answer in the present Letter is the physical mechanisms behind these
two terms.  To simplify the analyses we here focus on the behavior at the jamming density
$\phi=\phi_J$. An alternative, which is to examine the simulations in the hard disk limit,
is shown in Sec.~III~E of \REF{jointPRE}.  At $\phi=\phi_J$ \Eq{sigma-scale} simplifies
to,
\begin{equation}
  \label{eq:sigma-at-phiJ}
  \sigma(\phi_J,\gdot) = a_1\gdot^q + a_2\gdot^{q_2}.
\end{equation}
With $\phi_J\approx0.8434$ \cite{Heussinger_Barrat:2009, Olsson_Teitel:gdot-scale} and
with the methods described in \REF{jointPRE} we obtain the exponents $q=0.284(2)$ and
$q_2=0.567(7)$, in good agreement with previous works \cite{Olsson_Teitel:gdot-scale}. The
prefactors are $a_1=0.00437$ and $a_2=0.067$. The fit is shown in \Fig{sigma}(b).

\begin{figure}
  \includegraphics[bb=36 324 370 580, height=4.0cm]{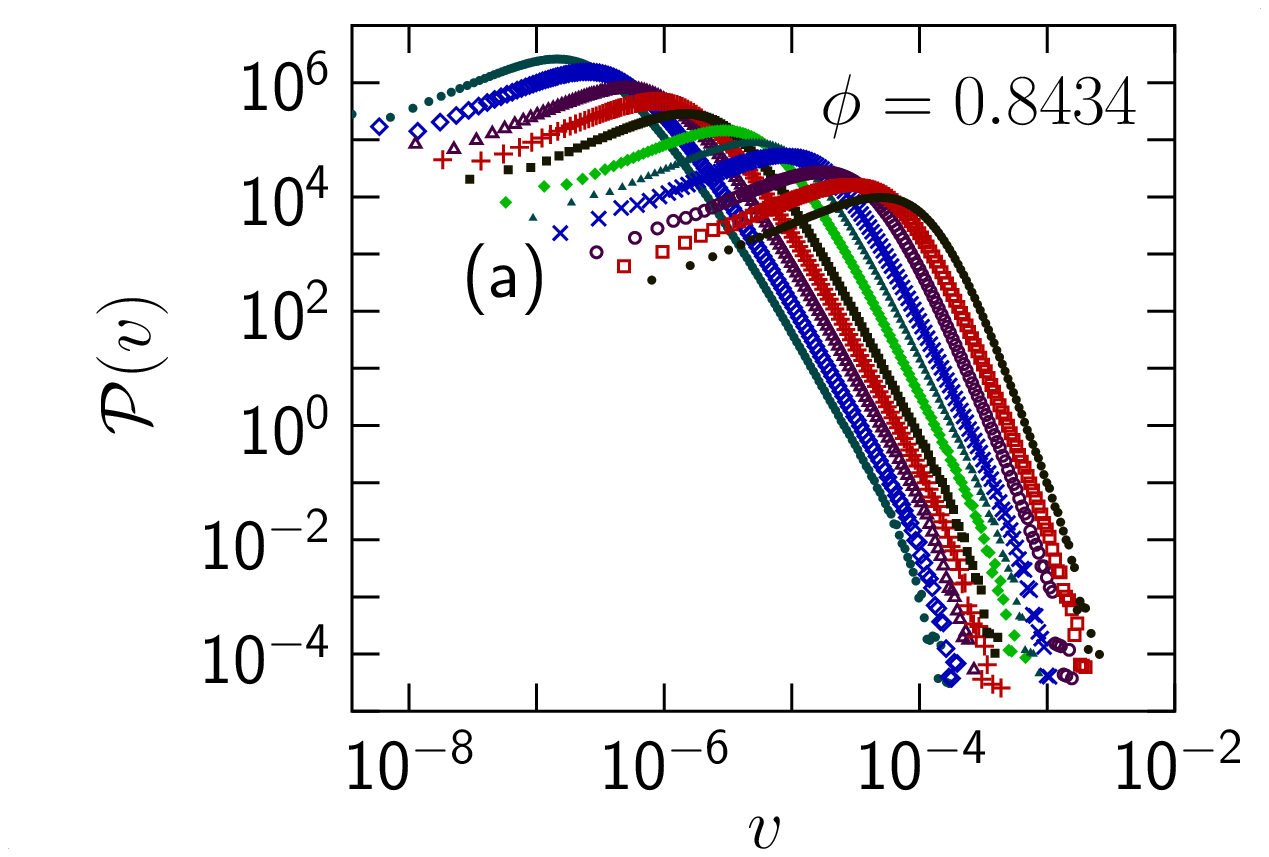}
  \includegraphics[bb=46 314 295 625, height=4.0cm]{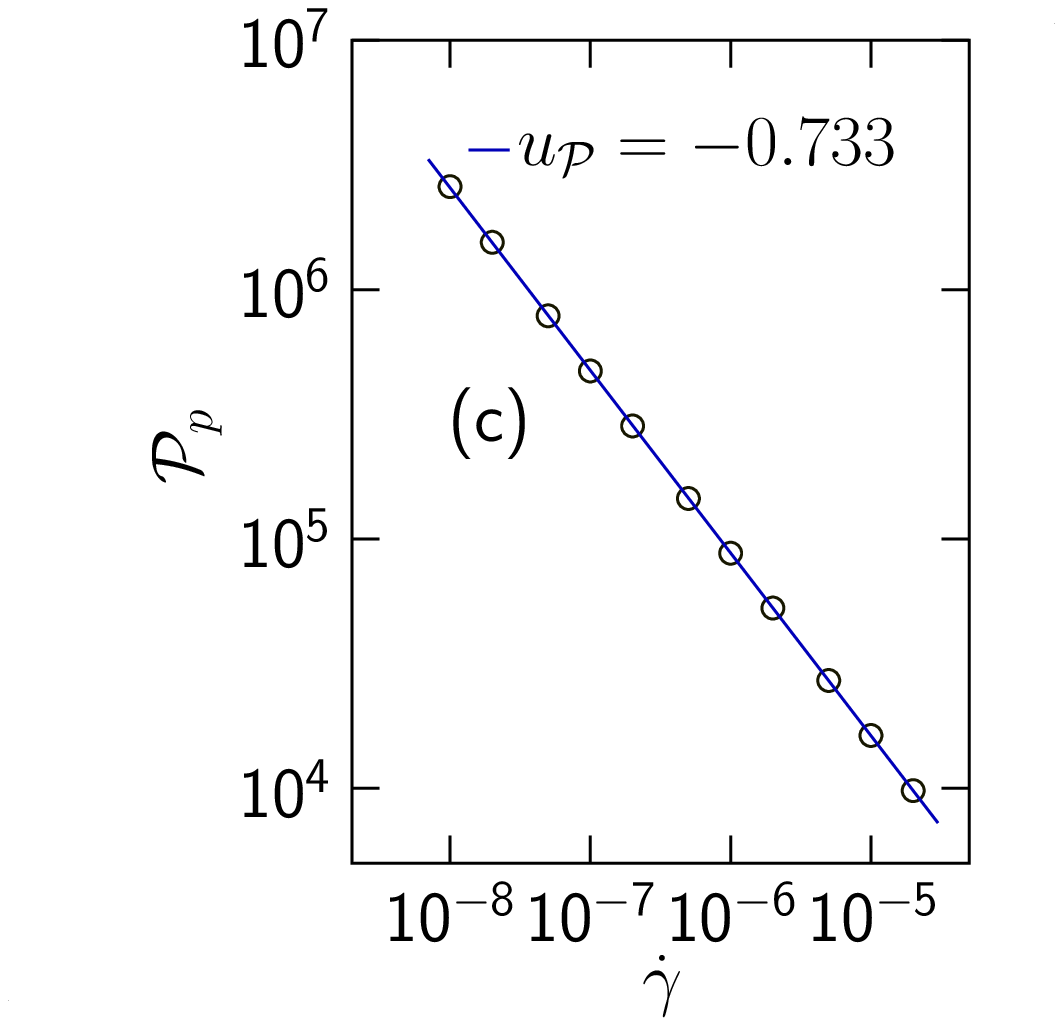}
  \includegraphics[bb=36 324 370 580, height=4.0cm]{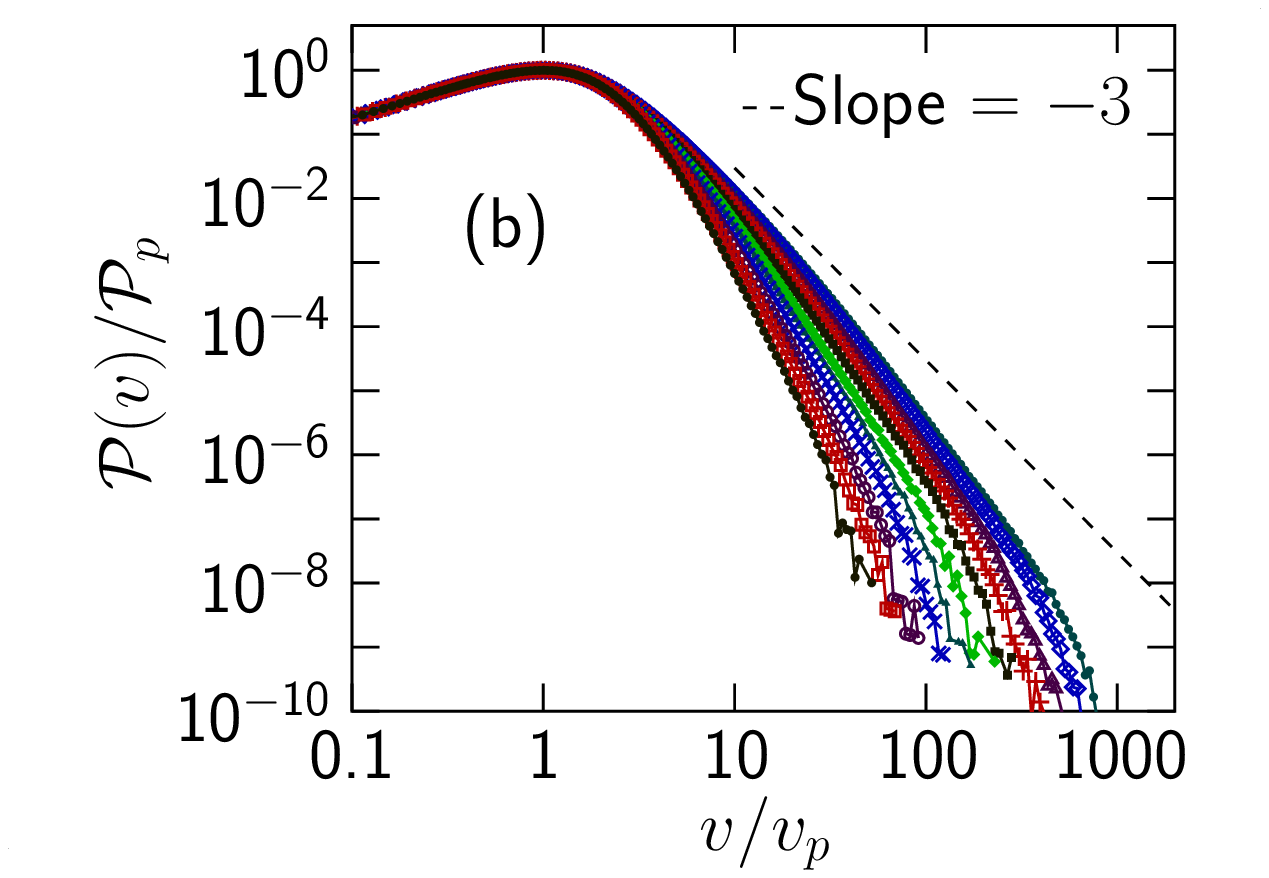}
  \includegraphics[bb=46 314 295 625, height=4.0cm]{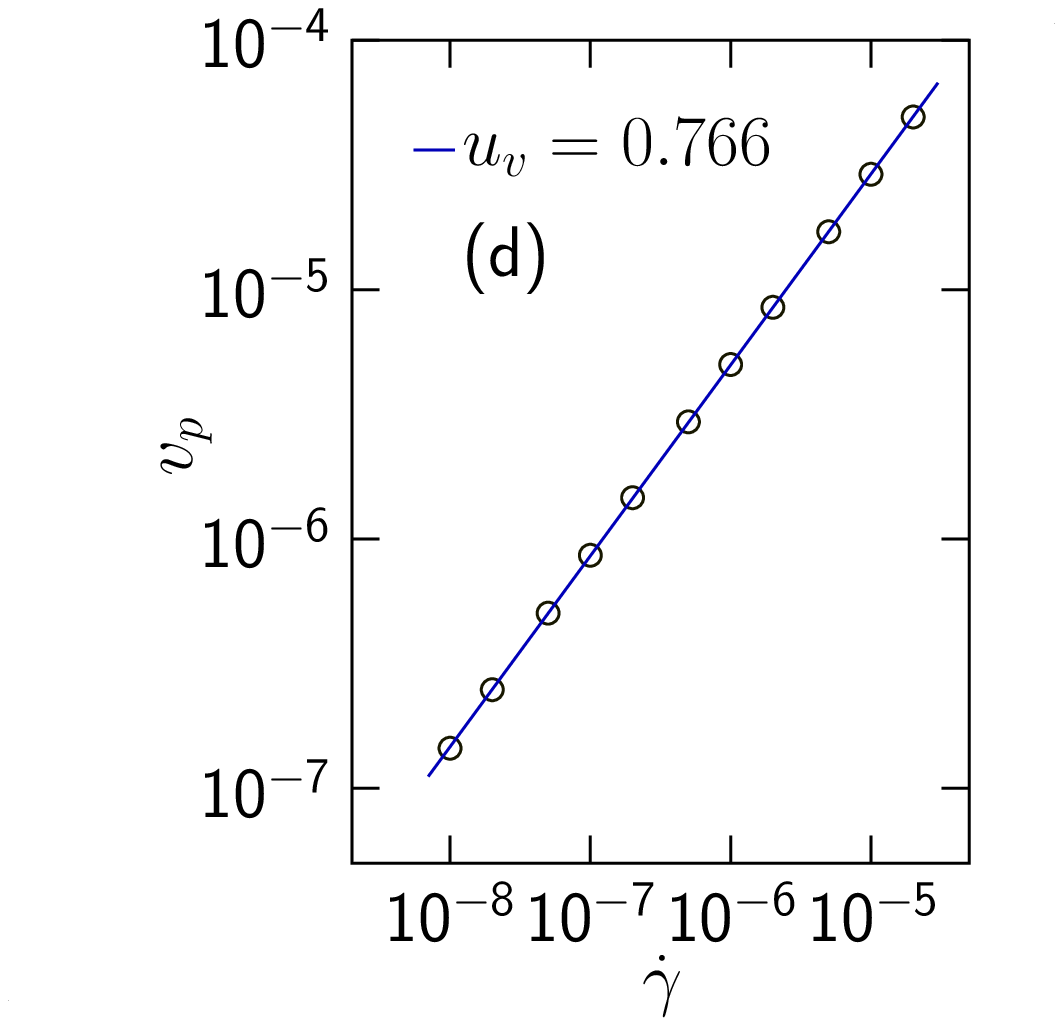}
  \caption{Velocity distributions at $\phi=0.8434\approx\phi_J$. Panel (a) gives $\cP(v)$
    for several different shear strain rates with symbols as in \Fig{sigma}(a).  Each data
    set has a clear peak and panel (b) shows the same data rescaled to make the peaks
    coincide. It is then found that the rescaled $\cP(v)$ collapse below and up to the
    peak whereas the data above the peak depend strongly on $\gdot$. At high velocities
    the distributions decay algebraically with a $\gdot$-dependent exponent. Panels (c)
    and (d) show the algebraic dependences of peak height and position on $\gdot$:
    $\cP_p\sim\gdot^{u_\cP}$ and $v_p\sim\gdot^{u_v}$.}
  \label{fig:vhist-v}
\end{figure}

To examine the shear stress in a novel way we turn to the properties of the velocity
distribution $\cP(v)$ calculated from the non-affine velocities $v_i\equiv|\v_i|$,
normalized such that $\int \cP(v) dv =1$. \Fig{vhist-v}(a) shows $\cP(v)$ vs $v$ at
$\phi=0.8434\approx\phi_J$ for $\gdot=10^{-8}$ through $2\times10^{-5}$. For each shear
strain rate there is a peak at $v=v_p$ with peak height $\cP_p=\cP(v_p)$.  A key
observation is now that the data up to and slightly above the peak collapse to a common
function, as shown in \Fig{vhist-v}(b), when $\cP(v)$ and $v$ are rescaled to make the
peaks fall on top of each other. The same is true also for $\cP(v)$ from a range of
densities both below and above $\phi_J$ (see \cite{jointPRE}). At higher velocities
$\cP(v)$ is algebraic, $\cP(v)\sim v^{-r}$, with an exponent $r$ that varies with $\phi$
and $\gdot$. [Earlier analyses suggest that $r\to3$ as jamming is approached
\cite{Olsson:jam-vhist}. The distributions are eventually cut off exponentially at large
$v$.]

We now set out to show that the second term in \Eq{sigma-at-phiJ} is related to the peak
in $\cP(v)$.  With the well known power balance $\sigma\gdot = (N/V) k_d \expt{v^2}$,
which is a relation between input power $V\sigma\gdot$ and dissipated power
$N k_d\expt{v^2}$ \cite{Ono_Tewari_Langer_Liu}, the shear stress may be written in terms
of $\cP(v)$ as
\begin{equation}
  \label{eq:sigma-v}
  \sigma = \frac N V \frac{k_d}{\gdot} \int \cP(v) v^2 dv.
\end{equation}
Introducing $x=v/v_p$ and $f(x)=\cP(v)/\cP_p$, the contribution to $\sigma$ from velocites
up to the peak becomes
\begin{equation}
  \label{eq:S-vp}
  S(v_p) = \frac N V k_d W_p \int_0^1 f(x) x^2 dx,
\end{equation}
where $W_p=\cP_p v_p^3/\gdot$. \Fig{vhist-v}(c) and (d) show that peak height and position
depend algebraically on $\gdot$: $\cP_p\sim\gdot^{u_{\cP}}$, with $u_{\cP}=-0.733$, and
$v_p\sim\gdot^{u_v}$, with $u_v=0.766$. For the $\gdot$-dependence of $S(v_p)\sim W_p$ we
then find
\begin{equation}
  \label{eq:Wp}
  W_p(\phi_J,\gdot) \sim \gdot^{u_{\cP}+3u_v-1}\sim \gdot^{u_w},
\end{equation}
with $u_w=0.565$, and note that this is in excellent agreement with
$q_2\approx 0.567$ from the fit of $\sigma(\phi,\gdot)$ to \Eq{sigma-at-phiJ}.

We now turn to the magnitude of the contribution due to the slow process. We formally
split $\cP(v)$ into two parts for the slow and the fast processes,
$\cP(v) = \cP_s(v) + \cP_f(v)$. With this splitting the dissipation from the slow process
becomes ${\cal D}_s=k_d\int \cP_s(v) v^2dv$. We get $\sigma_s=(N/V\gdot) {\cal D}_s$,
which means that $\sigma_s$ is determined from the dissipation.

To demonstrate that $\sigma_s$ from the dissipation is indeed equal to the
correction-to-scaling term $a_2\gdot^{q_2}$, in \Eq{sigma-at-phiJ}, we use the reasoning
behind \Eq{S-vp} with $f_s(x)\equiv\cP_s(v)/\cP_p$, to get
\begin{equation}
  \label{eq:sigmas}
  \sigma_s = \frac N V k_d W_p I_2,
\end{equation}
where $I_2\equiv\int_0^\infty f_s(x) x^2 dx$.  We find that the choice $I_2=3.4$ gives a good
agreement at $\phi_J$ between $\sigma_s$ and the second term of \Eq{sigma-at-phiJ}. In
\REF{jointPRE} it is shown that it is possible to construct a reasonable $f_s(x)$ that is
equal to $f(x)$ below the peak, decays exponentially at larger $v$, and gives
$I_2=3.4$. This shows that not only the $\gdot$-dependence, but also the magnitude, of the
correction-to-scaling term is consistent with it being due to the rescaled $f_s(x)$.

In Sec.~III\;E of \REF{jointPRE} we show this kind of analysis for data from the
hard disk limit at $\phi<\phi_J$ and show that it works well and gives consistent results
with the results here, through analyses from different $\gdot$ at $\phi_J$.

\begin{figure}
  \includegraphics[bb=70 338 413 700, width=3.0cm]{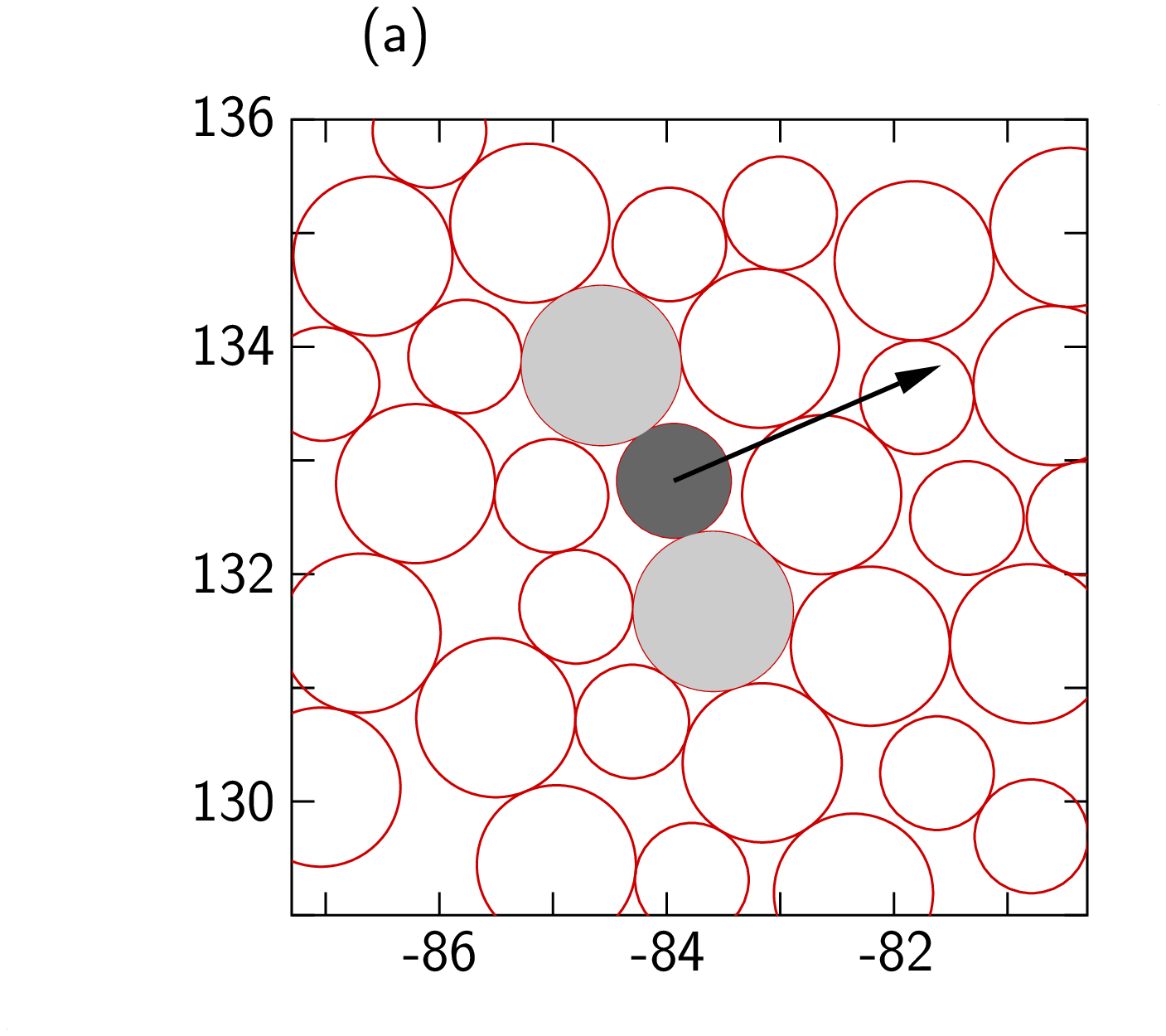}
  \includegraphics[bb=46 328 430 603, width=5.4cm]{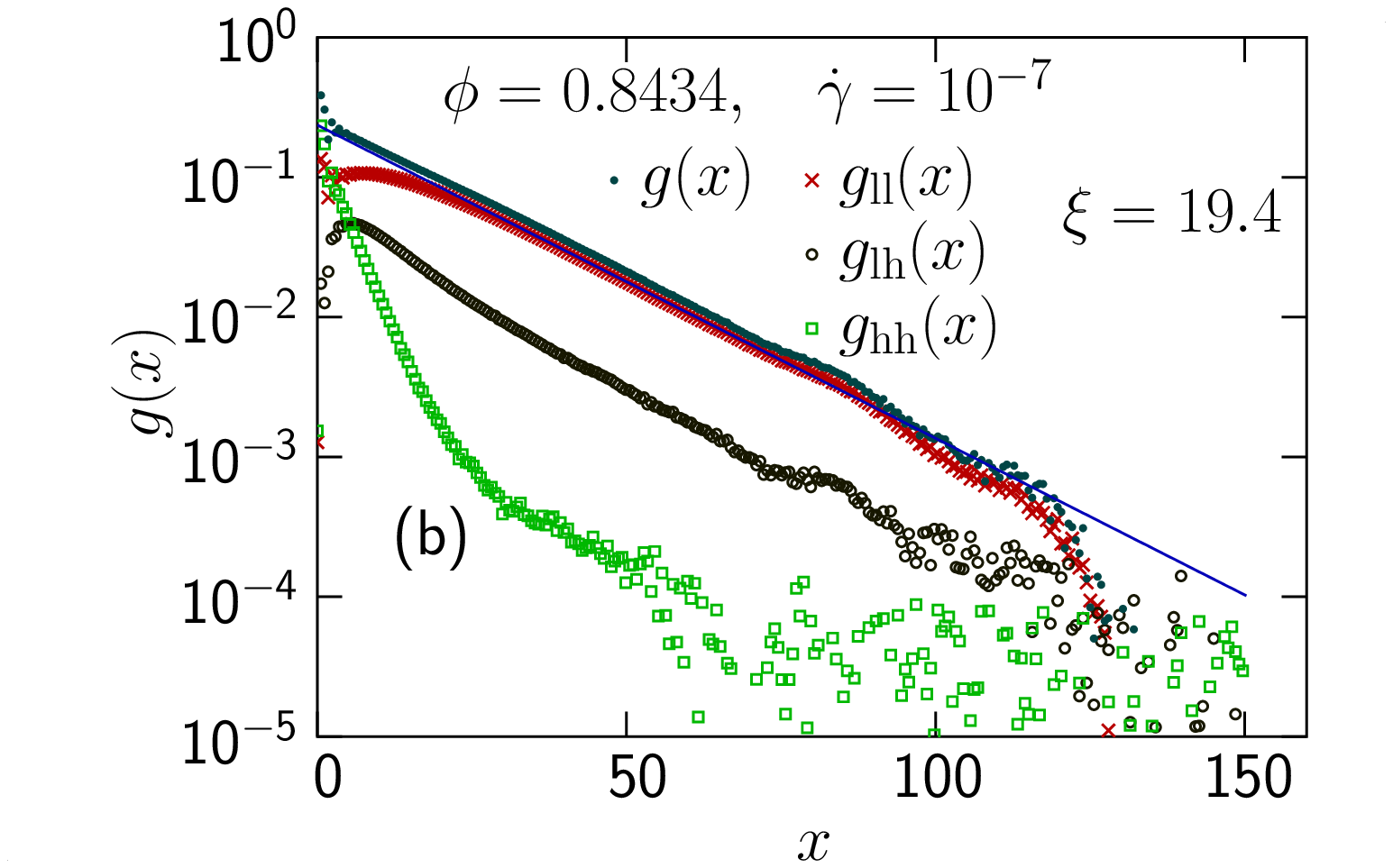}
  \caption{Behaviors of fast particles. Panel (a) which is a snapshot at $\phi=0.800$
    shows a particle with velocity $v/\expt v\approx 8.5$ which has a high velocity
    because it is not in a force-balanced state but is squeezed between two other
    particles, shown by light gray. Panel (b) shows the splitting of the velocity
    correlation function according to the low or high velocity of the colliding particles:
    $g_\mathrm{ll}$ for two low velocity particles, $g_\mathrm{hh}$ for two high velocity
    particles, and $g_\mathrm{lh}$ due to one particle with low and one with high
    velocity. The key message is that the contribution to the total $g(x)$ from two high
    velocity particles is very small.}
  \label{fig:snapshot}
\end{figure}

We now turn to analyzing the velocity correlations and then first note that the very wide
distribution of particle velocities is due to the fact that the particle velocity is
$v_i=f_i/k_d$, where the net force $f_i\equiv \sum_j f_{ij}$ is usually much smaller than
the typical contact force, $f_{ij}$, since the contact forces usually almost balance each
other out.  In cases where the contact forces fail to balance each other out, as e.g.\ in
\Fig{snapshot}(a) where the dark gray particle is squeezed between the two contacting
light gray particles, this may give an unusually large net force and thereby a high
velocity. The fastest particles are the ones with only two contacting particles but in the
Supplemental Material we show that they nevertheless give only about 7\% of the total
dissipation and that most of the dissipation is due to (short) chains of fast
particles. We also argue that the same mechanism that gives the fast particle in
\Fig{snapshot}(a) also gives these short chains of fast particles.

In \REF{Charbonneau:2015:prl} it was found that the theoretically expected force
distribution was obtained if contacts that were related to localized configurations were
not include in the calculation, and it was further argued that these contacts were due to
buckler configurations. We note the similarity of \Fig{snapshot}(a) and the buckler
configuration shown in \REF{Charbonneau:2015:prl} and remark that particles that are
irrelevant in the approach of \REF{Charbonneau:2015:prl} are here found to be
significant. \Fig{snapshot}(a) also suggests a possible relation between some of the fast
particles and irreversible contact changes in quasistatic shearing \cite{Morse:prr,
  Tuckman:SoftMatter:2020}.

It is well known \cite{Pouliquen:2004, Lechenault_2008, Heussinger_Berthier_Barrat:2010}
that the dynamics becomes increasingly collective as jamming is approached and a recent
paper has revealed a rich behavior of the velocity correlations
\cite{Olsson_Teitel:jam-xi-ell}. A large correlation length is what one would expect when
particles behave as a slowly moving fluid, but it is less clear what to expect for high
velocity particles, as in \Fig{snapshot}(a), which move erratically because of squeezing.
To answer this question \Fig{snapshot}(b) shows the velocity correlation
$g(x) = [\expt{v_\nearrow(0) v_\searrow(x\hat x)} + \expt{v_\searrow(0) v_\nearrow(x\hat
  x)}]/(\mathbf{v}^2/2)$ which is a measure of the rotation of the non-affine velocity
field \cite{Olsson_Teitel:jam-xi-ell}. We show both $g(x)$ from all particles and the
contributions to $g(x)$ from different sets of particles. We identify the velocities as
``low'' or ``high'' according to the threshold velocity $v_{50}$, chosen such that 50\% of
the dissipation is due to particles with $v<v_{50}$.
This is similar to the splitting into slow and fast processes, but also different since
$\cP_s(v)$ and $\cP_f(v)$ are overlaping, with no sharp threshold velocity.
\Fig{snapshot}(b) shows that the full $g(x)$ at $\phi=0.8434$ and $\gdot=10^{-7}$ decays
exponentially with a length
\label{fig:fast}$\xi\approx 19.4$ and that it is the low velocity particles that strongly
dominate the large distance correlation; $g_{hh}(x)$ from two high velocity particles,
contributes less than 1\% to the total $g(x)$, for large $x$.  We belive that this
non-zero value of $g_{hh}(x)$ is because the fast particles also get a contribution to their
velocities from the slow process since they are embedded in a set of particles
that behaves as a slowly moving fluid.

This therefore suggests that the large correlation length is due to the slow process,
only, and this is a finding with profound consequences as it suggests that one process is
responsible for the correlations whereas another process is behind the leading divergence
in the viscosity, which is at odds with usual critical phenomena.
The link between viscosity and correlation length would therefore seem to be
an indirect one, only, and it appears that there exists some, as yet unknown, mechanism
that connects the slow and the fast processes together.

\begin{figure}
  \includegraphics[bb=20 326 532 577, width=7cm]{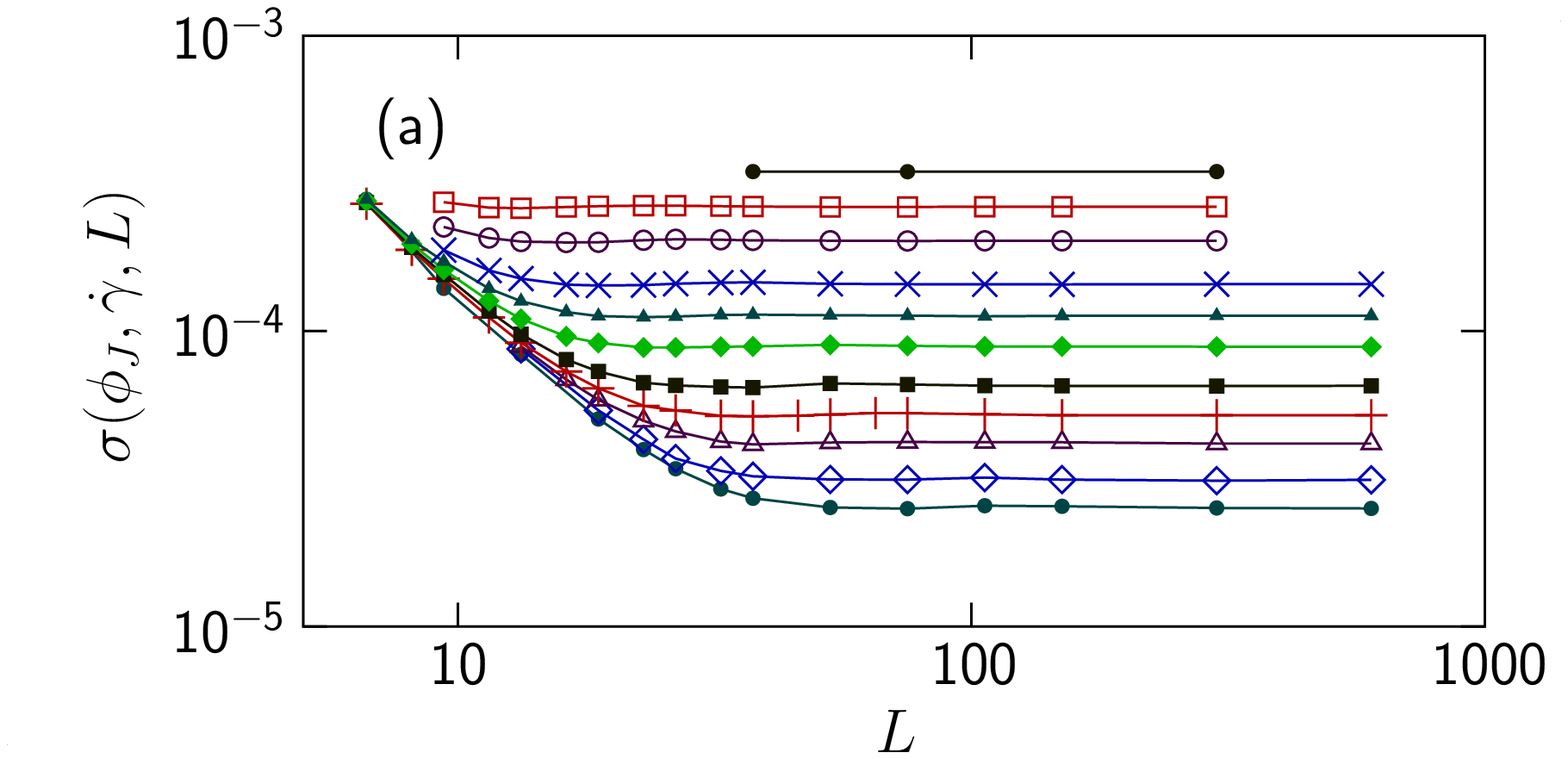}
  \includegraphics[bb=20 326 532 577, width=7cm]{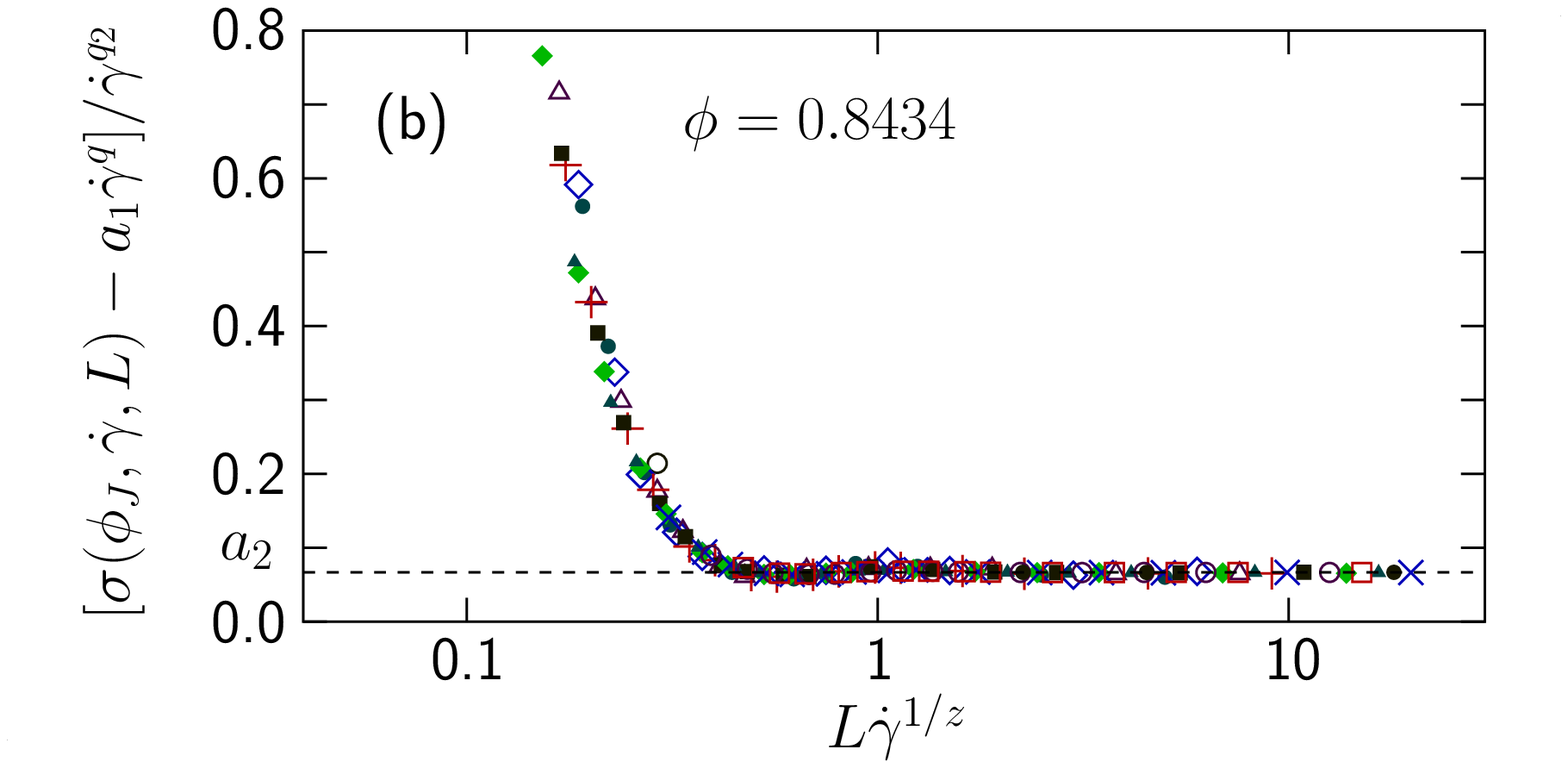}
  \caption{Finite size scaling at $\phi_J\approx 0.8434$. Panel (a) is the raw data
    $\sigma(\phi_J,\gdot,L)$ for $\gdot=10^{-8}$ through $2\times10^{-5}$, with symbols as
    in \Fig{sigma}(a), for $N=32$ through 262144. Panel (b) shows that
    $[\sigma-a_1\gdot^q]/\gdot^{q_2}$ collapses when plotted vs $L \gdot^{1/z}$, in
    agreement with \Eq{sigma-scaling-L-a1}. We here use $1/z=0.26$, assuming $1/z\nu=0.26$
    \cite{Olsson_Teitel:gdot-scale} and $\nu=1$ \cite{Olsson_Teitel:jam-xi-ell}.}
  \label{fig:sigmax-Lgdot}
\end{figure}

The lack of large distance correlations of the fast particles has consequences also for the
finite size dependence, examined in simulations with different $N\propto L^2$. To
include the $L$-dependence in the critical scaling analysis one adds $b/L$ as an
additional argument to the scaling functions of \Eq{sigma-b}, and \Eq{sigma-at-phiJ} then becomes
\begin{equation}
  \label{eq:sigma-scaling-L}
  \sigma(\phi_J, \gdot, L) = \gdot^q \tilde g_\sigma(L \gdot^{1/z}) + \gdot^{q_2} \tilde
  h_\sigma(L \gdot^{1/z}).
\end{equation}
This means that one would expect $\sigma(\phi_J, \gdot, L)$ to be the sum of two
$L$-dependent functions, respectively approaching $a_1$ and $a_2$ for large $L$, with
different prefactors, $\gdot^q$ and $\gdot^{q_2}$. It does however turn out that the data,
shown in \Fig{sigmax-Lgdot}(a), fit very well to a simpler expression without any finite
size dependence in the first term,
\begin{equation}
  \label{eq:sigma-scaling-L-a1}
  \sigma(\phi_J, \gdot, L) = a_1 \gdot^q + \gdot^{q_2} \tilde h_\sigma(L \gdot^{1/z}).
\end{equation}
This is demonstrated in \Fig{sigmax-Lgdot}(b) which shows that
$[\sigma(\phi_J, \gdot, L) - a_1 \gdot^q]/\gdot^{q_2}$, with $q$, $q_2$, and $a_1$ the
same as in \Eq{sigma-at-phiJ}, collapses onto a single function when plotted vs
$L\gdot^{1/z}$. Even though there is nothing in the formalism that excludes the
possibility that the function $\tilde g_\sigma(L\gdot^{1/z})$ could be a constant $=a_1$,
the absence of a clear finite size dependence for the leading diverging term is clearly at
odds with the ordinary behavior in critical phenomena.  It is however entirely in
accordance with the absence of correlations across large distances in \Fig{snapshot}(b)
for the particles with higher velocities, which are the ones that dominate the
$\gdot^q$-term.

To summarize, we provide strong evidence that shear-driven jamming is governed by two
processes with different properties: The fast process is responsible for the leading
divergence of the shear viscosity whereas the slow process is behind the diverging
correlation length. The absence of a direct coupling between these diverging quantities
suggests that shear-driven jamming is an unusual kind of critical phenomenon.

\begin{acknowledgments}
  I thank S. Teitel for many discussions and suggestions on the manuscript. The
  computations were enabled by resources provided by the Swedish National Infrastructure
  for Computing (SNIC) at High Performance Computer Center North, partially funded by the
  Swedish Research Council through grant agreement no.\ 2018-05973.
\end{acknowledgments}

%

\end{document}